# A Unified Phonon Interpretation for the Non-Fourier Heat Conduction by Non-equilibrium Molecular Dynamics Simulations

Yue Hu[1], Xiaokun Gu[2], Tianli Feng[3], Zheyong Fan[4], Hua Bao[1,*]

[1]University of Michigan-Shanghai Jiao Tong University Joint Institute, Shanghai Jiao Tong University, Shanghai 200240, P. R. China

[2]Institute of Engineering Thermophysics, School of Mechanical Engineering, Shanghai Jiao Tong University, Shanghai 200240, China

[3]Materials Science and Technology Division, Oak Ridge National Laboratory, Oak Ridge, Tennessee 37831, USA.

[4]QTF Centre of Excellence, Department of Applied Physics, Aalto University, FI-00076 Aalto, Finland

**Abstract**

Nanoconfinement induces many intriguing non-Fourier heat conduction phenomena that have been extensively studied in recent years, such as the nonlinear temperature profile inside the devices, the temperature jumps near the contacts, and the finite-size effects. The understanding of these phenomena, however, has been a matter of debate over the past two decades. In this work, we demonstrate a unified phonon interpretation of non-Fourier heat conduction which can help to understand these phenomena by a mode-to-mode correspondence between the non-equilibrium molecular dynamics (NEMD) simulations and the mode-resolved phonon Boltzmann transport equation (BTE). It is found that the nanoscale phonon transport characteristics including temperature profile, the heat flux value and the modal temperature depend on the applied thermal reservoirs on the two contacts. Our NEMD simulations demonstrate that Langevin thermostat behaves like an infinitely large thermal reservoir and provides thermally equilibrium mode-resolved phonon outlets, while biased reservoirs, e.g., Nose-Hoover chain thermostat and velocity rescaling method behave like non-equilibrium phonon outlets. Our interpretation clearly demonstrates that the non-Fourier heat transport phenomena are originated from a combination of non-diffusive phonon transport and phonon thermal nonequilibrium. This work provides a clear understanding of nanoscale heat transport and may guide the measurement and control of thermal transport in various applications.

---

[*] Author to whom correspondence should be addressed. *E-mail address*: hua.bao@sjtu.edu.cn (H.Bao).

I. Introduction

Nanoscale heat transport is critical for thermal management of electronics and thermoelectric energy harvesting[1,2]. Non-equilibrium molecular dynamics (NEMD) simulation is an effective method to study nanoscale heat conduction, and it has been used in many systems[3–12]. The implementation of NEMD simulation is analogous to steady state experimental setup in thermal conductivity measurement, in which two reservoirs (one heat source and one heat sink) are added to the system to generate a 1D steady state heat transfer in the sample region. To generate the reservoirs, one can either apply a constant heat flux by the velocity rescaling (VR) method[13,14] or a temperature difference by thermostats such as the Langevin thermostat[15] and the Nose-Hoover chain (NHC) thermostat[16–18]. By measuring the ratio of the heat flux and the temperature gradient, the thermal conductivity of the sample can be obtained according to the Fourier law. Although the idea is straightforward, some unexpected non-Fourier phenomena appear in NEMD simulations, particularly, the nonlinearity of the temperature profile in the sample region, the temperature jump between the heat source/sink and the sample, and the finite-size effects of the predicted thermal conductivity. The major open questions are: 1) whether these phenomena have physical interpretations; 2) how to treat these effects in the simulations to reliably obtain the transport properties of the system under study.

Over the past two decades, there are quite many discussions on these non-Fourier phenomena observed in NEMD simulations. The first phenomenon is the nonlinearity of the temperature profile, i.e., the temperature gradient is not constant when the heat flux is constant. This phenomenon is usually observed when the sample is a good thermal conductor[19,20]. A natural question is whether the Fourier law is valid in this nonlinear region. Howell and Lukes *et al.* showed that the velocity distribution of atoms follows the Boltzmann distribution, indicating that the time-averaged velocities reach thermal equilibrium and the Fourier law is vaild[19,21]. Thus, the reason for the nonlinearity was sometimes understood as the additional strong scattering near the reservoirs[19,20,22]. To obtain a well-defined thermal conductivity, a common strategy is to use the linear-fit method, i.e. exclude the nonlinear part near the reservoirs and extract the temperature gradient from a near-linear part of the temperature profile. The second phenomenon is the temperature jump between the heat source/sink and the sample that is only observed in NEMD simulations using thermostats, and not in NEMD simulations using the VR method[23]. The temperature jump was regarded as a numerical artifact related to some simulation parameters[15,24]. Dhar suggested that the temperature jump is caused by the contact resistance which should be reduced by using an optimal value of the coupling parameter of the thermostat[15]. Chen and Li studied the effects of thermostats and also suggested that the temperature jump should be reduced by modifying the coupling parameter and increasing the length of the thermostatted regions[24]. The third phenomenon is the finite-size effects, i.e. the thermal conductivity increases when the length of the simulation cell increases. This phenomenon arises when the length of the simulation cell is not significantly longer than the mean free path[20,25]. The common understanding of the finite-size effects is that ballistic phonons encounter scatterings at the boundaries between the sample and the reservoirs. In order to solve this issue, Sellan *et al.* derived a model based on the phonon Boltzmann transport equation (BTE) and the Matthiessen rule[22]. Based on this model, an extrapolation method can be used to get the bulk thermal conductivity from several size-dependent thermal conductivities[26]. To simplify the complexity, the linear extrapolation is commonly used[26,27].

The above conclusions have been widely adopted in NEMD simulations[28]. However, new insight related to these phenomena has been provided in more recent investigations. For example, Feng *et al.* developed a spectral phonon temperature (SPT) method and discovered the local spectral phonon thermal non-equilibrium in their cases directly[29]. Thus, the Fourier law is not valid in these cases. Therefore, the physical meaning of using the linear-fit method to calculate the thermal conductivity is still unclear. Recently, by comparing the NEMD results with results from atomistic Green's function[30] (AGF) and homogenous non-equilibrium molecular dynamics[31] (HNEMD) (a method similar to equilibrium molecular dynamics), Li *et al.* concluded that the temperature jump has physical reason and should be included in thermal conductivity calculation[32]. Also, different size-dependent thermal conductivities are obtained by using the Langevin thermostat and the NHC thermostat[32,33], indicating that the previous extrapolation model may not be able to describe the finite-size effects for all kinds of reservoirs.

Recent findings clearly show that the previous understandings of non-Fourier heat conduction by NEMD simulations are insufficient. This insufficiency may lead to the large discrepancy of the results between NEMD and other simulation methods or experiments even for the same system in some previous works. Previous investigations rarely discussed the origins of the differences among using different reservoirs in NEMD. Also, because NEMD simulations study random lattice vibration in real space, it is difficult to directly extract the phonon physics. In contrast, the phonon gas model has been widely adopted to understand the ballistic to diffusive thermal transport[19,22,23,33]. In this work, we investigate the thermal transport of silicon confined between two thermal reservoirs, using both NEMD and solution of phonon BTE. In NEMD, we use the Tersoff potential to describe the interatomic interactions. In BTE, we extract the phonon modal information with the same classical potential. Then, a direct comparison between NEMD and phonon BTE is performed to extract the phonon picture behind NEMD studies. As will be shown, with correct modeling, a quantitative agreement between NEMD and phonon BTE can be achieved. This manuscript is organized as follows. In Sec. II, NEMD simulations with sufficiently large thermal reservoirs will be carried out. We will revisit the difference among the Langevin thermostat, the NHC thermostat, and the VR method. In Sec. III, quantitative comparison and a mode-to-mode correspondence between NEMD and mode-resolved phonon BTE will be provided, from which phonon interpretations of the non-Fourier heat conduction can be obtained. In Sec. IV, some important issues in the phonon interpretations are further proved by NEMD simulations. In Sec. V, we will discuss the above phenomena according to the phonon interpretations. In Sec.VI, we will give a summary and conclusions.

## II.  NEMD simulations

An experimental thermal conductivity measurement setup example is shown in Fig. 1, in which a sample is attached to a heat source and heat sink. By measuring the heat flux and temperature difference, one can obtain the thermal conductivity by the Fourier's law.

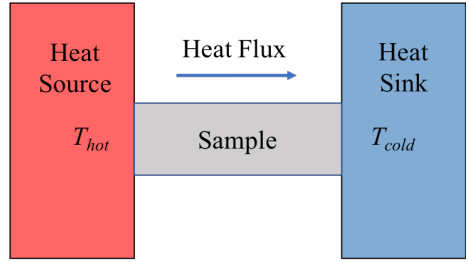

Fig. 1. Schematic illustration of thermal conductivity measurement. A sample is attached to a heat source and heat sink. One can either apply a temperature gradient in the sample by setting up a temperature difference and measure the resulting heat flux, or add a heat flux and measure the resulting temperature gradient inside the sample.

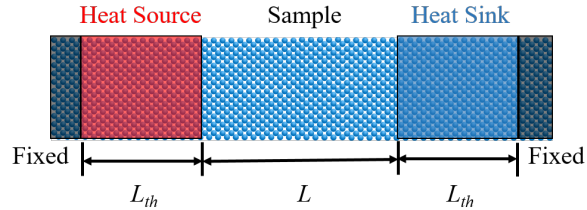

Fig. 2. Schematic illustration of the simulation cell used in NEMD simulations. The simulation cell is a piece of atomic structure. Two thermal reservoirs (one heat source and one heat sink) with a length of $L_{th}$ are established. The region between two thermal reservoirs is defined as the sample region with a length of $L$. A few layers of atoms at the two ends of the transport direction are fixed. Periodic boundary conditions are set in the directions perpendicular to the transport direction.

In NEMD simulations, the system is modeled using the atomic structure as shown in Fig. 2. Two thermal reservoirs (one heat source and one heat sink) with a length of $L_{th}$ are established by depositing/extracting energy from the atoms inside the reservoirs. The region between the two thermal reservoirs is defined as the sample region with a length of $L$. To prevent the atoms in the thermal reservoirs from sublimating, a few layers of atoms at the two ends of the transport direction are fixed. Periodic boundary conditions are set in the directions perpendicular to the transport direction. There are three representative thermal reservoirs, including the VR method[13,14], which controls the heat flux, the Langevin thermostat[15] and the NHC thermostat[16,17], which control the temperature.

We choose silicon modeled by the Tersoff potential[34] as the testing material throughout this work. All the NEMD simulations were implemented in the Large-scale Atomic/Molecular Massively Parallel Simulator (LAMMPS) package[35]. The cross-sectional area is set as that of 8×8 unit cells, which are large enough to eliminate the finite-size effects in the transverse directions. A time step of 1 fs, which ensures good energy conservation, was used in all the MD simulations. To obtain temperature and heat flux profiles, we first equilibrated the whole simulation cell under the NPT ensemble for 5 ns, after which we switched off the NPT ensemble and established two thermal reservoirs. The two thermal reservoirs are established for 20 ns. The data within the last 10 ns are used to extract temperature and heat flux. The heat flux is calculated from the stress formula in LAMMPS, which is problematic for the Tersoff potential[36,37]. However, it can result in comparable results with the correct formula for silicon crystal[38]. The heat flux values in the sample are also close to those obtained by energy conservation[20] in our cases. Thus, we still use the stress formula in the LAMMPS for simplicity. We also use the SPT method developed by Feng *et al.*[29,39] to extract the phonon modal temperature in the NEMD simulations. For the cases that used the SPT method, we first equilibrated the whole simulation cell under the NPT ensemble for 10 ns, after which we switched off the NPT ensemble and established two thermal reservoirs. The two thermal reservoirs are established for 40 ns. The data within the last 20 ns are used in the SPT method.

NEMD simulations are conducted by using the three methods mentioned above. For all the three methods, the length of the thermal reservoir is first set as a relatively large value of 25 nm, and later its size effect will be discussed. To consider the size effect of the simulation domain, two lengths, 13 nm and 56 nm, are used for the sample region. Since they are both smaller than the phonon mean free path of silicon, finite-size effects should exist[22]. The time parameter is set as 0.1 ps for the Langevin thermostat and the NHC thermostat, which is recommended by a previous study[32]. The target temperature is 310 K in the heat source and 290 K in the heat sink for the Langevin thermostat and the NHC thermostat. The amount of heat added for every time step is 2.50 meV for the 13 nm case and 1.66 meV for 56 nm case in the VR method.

The temperature and heat flux profiles are shown in Fig. 3. For both lengths, the temperature profiles using the Langevin thermostat are distinct from those using the NHC thermostat and the VR method. For the Langevin thermostat, we can see that the temperature is constant inside the thermal reservoirs except for a small region near the sample, while the temperature varies at different positions inside the thermal reservoirs for the NHC thermostat and the VR method. Temperature jumps can be observed near the boundary of the thermal reservoirs for the Langevin thermostat but not for the NHC thermostat and the VR method. For all the cases, nonlinearity exists in the sample region. The Langevin thermostat seems to give a smaller slope compared to the NHC thermostat and the VR method.

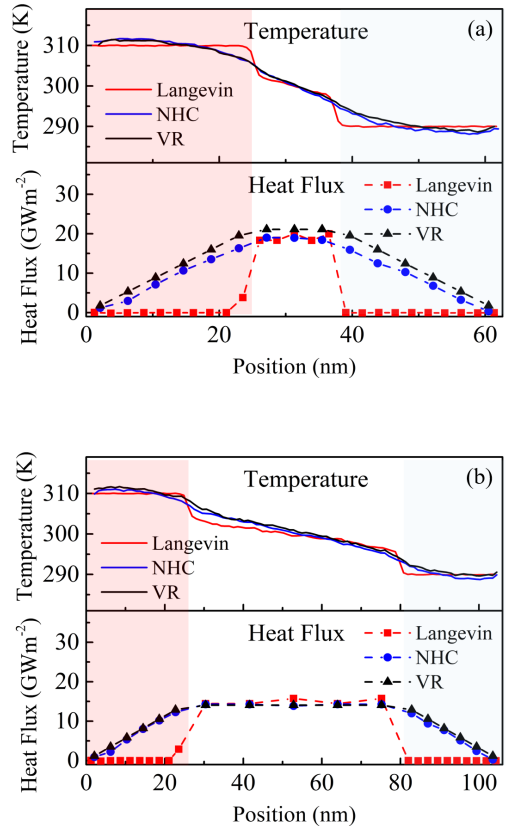

Fig. 3. Temperature and heat flux profiles for silicon with different lengths: (a) 13nm, (b) 56nm by using the Langevin thermostat (Langevin), the NHC thermostat (NHC), and the VR method (VR). The shaded regions represent the heat source (red) and heat sink (blue), and the region in between represents the sample.

The heat fluxes are also plotted in the bottom of Fig. 3(a) and (b). For the Langevin thermostat, the value of the heat flux is zero inside the thermal reservoirs except for a small region near the sample. In contrast, the heat flux is linearly increasing away from the adiabatic boundary inside the thermal reservoirs for the NHC thermostat and the VR method. According to energy balance ($\int \dot{Q} dV = \int \mathbf{q} \cdot \mathbf{n} dA$, where $\dot{Q}$ is the volumetric heat generation rate, $\mathbf{q}$ is the heat flux and $\mathbf{n}$ is the surface normal), which should be always valid, this means that the heat generation inside the thermal reservoirs is zero for the Langevin thermostat, while is nearly uniform for the NHC thermostat and the VR method. As such, at steady state, the Langevin thermostat only deposits heat in a small region near the sample and keeps the temperature at the target for every position inside the thermal reservoirs (except for the small region near the sample). In the NHC thermostat and the VR method, uniform heat generation/extraction is applied at different positions inside the thermal reservoirs.

To further examine the modal temperature of different phonon modes, the SPT method[29,39] was applied to analyze the simulation data. For simplicity, it is only applied for the case of $L = 13$ nm. We calculate the phonon temperature for 204 phonon modes and plot the average value for 6 different phonon branches in Fig. 4.

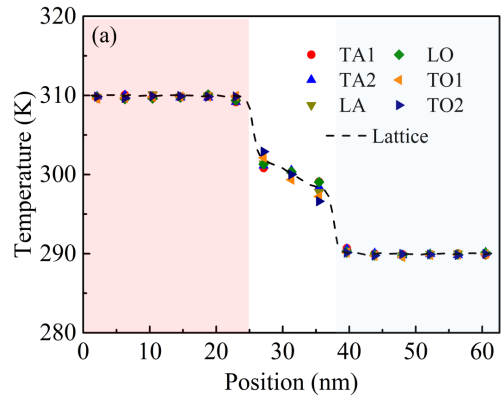

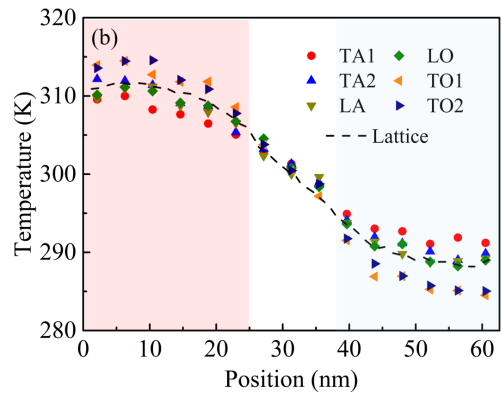

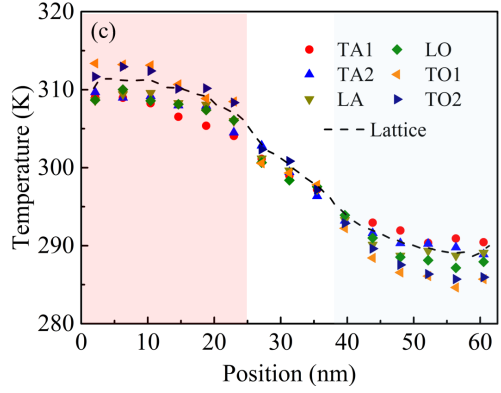

Fig. 4. Averaged temperature profiles for phonon modes in 6 different branches from NEMD simulation with (a) the Langevin thermostat, (b) the NHC thermostat, and (c) the VR method.

Again, we observe distinct behavior for the Langevin thermostat, while the NHC thermostat and the VR method behave similarly. For the Langevin thermostat, the modal temperatures are not at equilibrium in the sample region but are almost at equilibrium inside the thermal reservoirs except for a small region near the sample. For the NHC thermostat and the VR method, the modal temperatures are strongly out of equilibrium not only in the sample region but also inside the thermal reservoirs. We note that Dunn *et al.* obtain the opposite results to us by detecting the spectral energy distribution in the reservoirs. They found phonon nonequilibrium in the reservoirs for the Langevin thermostat but not found that for the NHC thermostat[33]. The reason is that they perform the spectral analysis by sampling the whole reservoir and thus local nonequilibrium information is averaged out.

From these results, we can distinguish different nanoscale phonon transport behaviors including temperature profile, heat flux value and model temperature by using different thermal reservoirs. The Langevin thermostat behaves differently, while the NHC thermostat and the VR method are similar to each other. As such, in the subsequent discussions, only the Langevin and the NHC thermostats are considered. The conclusions of the NHC thermostat should apply to the VR method as well.

### III.  Phonon BTE analysis

Based on the NEMD results from Sec. II, we notice different behaviors in the NEMD simulations with different reservoirs. In this section, we use the mode-resolved phonon BTE to model the systems as studied in the NEMD simulations. From Fig. 3, we note that for the Langevin thermostat, the temperature is uniform inside the thermal reservoirs, which is similar to an infinitely large constant temperature thermal reservoir. Therefore, we consider the thermalizing boundary condition in the BTE calculations, as shown in Fig. 5(a). For the NHC thermostat, since the heat generation inside the thermal reservoir is uniform, we use a uniform heat generation, as shown in Fig. 5(b). The fixed atoms do not exchange energy with the atoms in the thermal reservoir, and thus constitute an adiabatic boundary. In the framework of phonon BTE, the adiabatic boundary can be specular or diffuse, or a mixture of both[40]. Here we temporarily consider specular boundary, and will discuss this effect in details later.

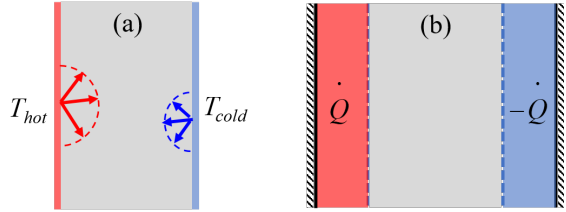

Fig. 5. The 1D simulation domain of phonon BTE with (a) thermalizing boundary condition and (b) uniform heat generation and adiabatic boundary conduction.

The thermalizing boundary condition in phonon BTE behaves like a black surface with the target temperature, in which we set as 310 K for $T_{hot}$ and 290 K for $T_{cold}$, to be consistent with the NEMD simulations. This surface emits outgoing phonons with the target temperature into the sample region, while all the incoming phonons into the surface are absorbed. For uniform heat generation in Fig. 5(b), uniform thermal energy is added to the heat source and the same amount is extracted from the heat sink. Since our BTE simulation is mode-resolved[41], the heat should be added to every phonon mode. It is difficult to obtain the amount of energy added to each phonon mode in the NEMD simulations. Therefore, the energy added to each phonon mode is simply chosen to be proportional to the heat capacity of the mode. The lengths of the thermal reservoirs and the sample are set to the same as in the NEMD simulations.

We adopted the finite volume method[42] to numerically solve the mode-resolved phonon BTE under the relaxation time approximation[43]. To solve the mode-resolved phonon BTE, one needs the group velocity $\mathbf{v}_{\omega,p}$, the relaxation time $\tau_{\omega,p}$ and the heat capacity $C_{\omega,p}$ for every phonon mode as the input information. We emphasize that the input information is extracted from the same system as NEMD: silicon crystal with Tersoff potential at 300 K. These parameters are obtained using the standard anharmonic lattice dynamics approach[44,45], in which the harmonic and anharmonic interatomic force constants are first extracted by fitting the relation between atomic forces, $F$, and the displacements, $u$,

$$F_i^\alpha = -\sum_j \sum_\beta \phi_{ij}^{\alpha\beta} u_j^\beta - \frac{1}{2!}\sum_{jk}\sum_{\beta\gamma}\psi_{ijk}^{\alpha\beta\gamma} u_j^\beta u_k^\gamma \ldots$$

,where $(i,\alpha)$ means the $\alpha$ direction of atom $i$, $\phi$ and $\psi$ are the harmonic and third-order anharmonic force constants. The phonon dispersion is obtained by diagonalizing the dynamical matrix produced by the harmonic force constants. With the dispersion relation $\omega(\mathbf{q},s)$, the group velocity and heat capacity of the mode $(\mathbf{q},s)$ are simply calculated through $\mathbf{v}(\mathbf{q},s) = \partial\omega(\mathbf{q},s)/\partial\mathbf{q}$ and $c(\mathbf{q},s) = \partial\hbar\omega(\mathbf{q},s)n^0(\mathbf{q},s)/\partial T$ with the phonon population function being $n^0(\mathbf{q},s)$. In order to fairly compare the NEMD and phonon BTE results, the phonon population function used in this work has the same form as the standard Bose-Einstein distribution, $n^0(\mathbf{q},s) = 1/\left[\exp(\hbar)\omega(\mathbf{q},s)/k_B T\right]$, but with a modified Planck's constant, which is 1/100 of the original value[46]. This treatment could reproduce the classical distribution in MD simulations. We compute the phonon relaxation time through the lowest-order perturbation theory, in which three-phonon processes are regarded as the only source for phonon-phonon scatterings. The computation of the three-phonon relaxation times requires the third-order anharmonic force constants and the expression can be found in previous publications[47,48]. With these phonon properties, the thermal conductivity of silicon is calculated based on the single mode relaxation time approximation method. We use 10×10×10 q-points to sample the Brillouin zone. The obtained classical thermal conductivity is 245 W/mK at 300 K, which is very close to the previous EMD result for Tersoff silicon[26]. In the BTE solver, we are not able to consider as many phonon modes as in the system due to the huge computational cost, so we use the information from 100 phonon bands obtained by averaging over the different modes. The details of averaging can be found in the previous publications[41,49].

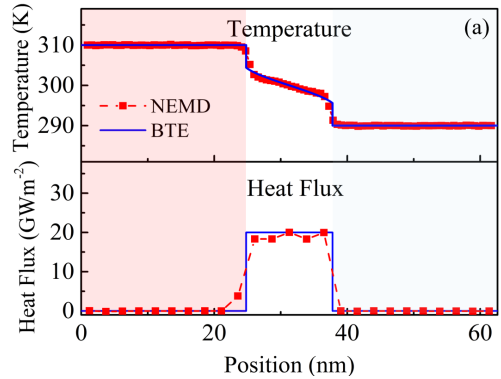

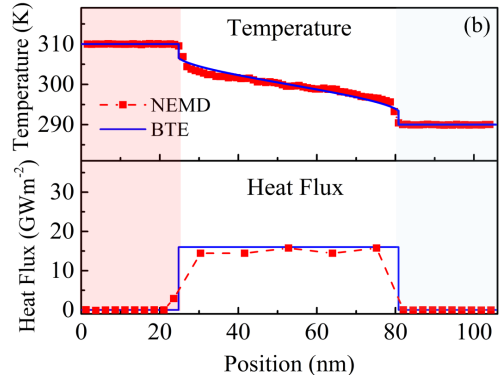

Fig. 6. Comparisons of temperature profiles and heat flux profiles between the NEMD simulations with the Langevin thermostat (NEMD) and the BTE calculations with the thermalizing boundary condition (BTE) for two different sample lengths of (a) 13.0 nm and (b) 56.0 nm.

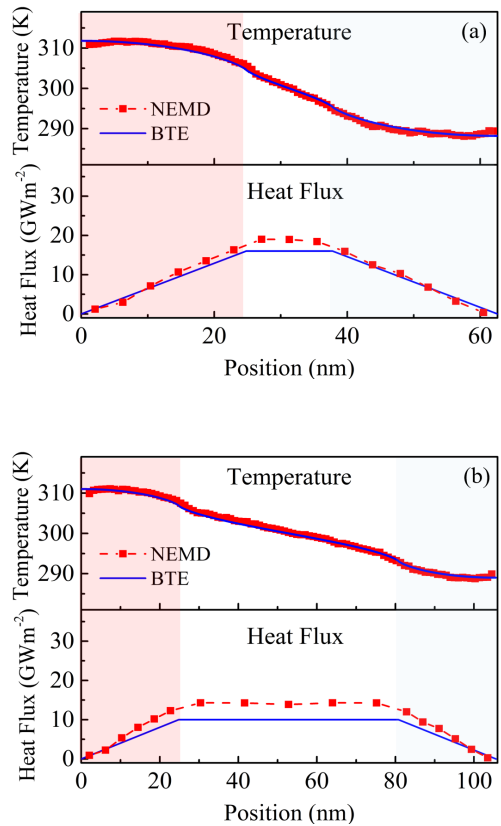

Fig. 7. Comparisons of temperature profiles and heat flux profiles between NEMD simulations with the NHC thermostat (NEMD) and the BTE calculations with uniform heat generation (BTE) for two different sample lengths of (a) 13nm and (b) 56nm.

The temperature and heat flux profiles, obtained from the solution of the BTE, are plotted together with the NEMD results, as shown in Fig. 6 and Fig. 7. From Fig. 6(a) and (b), we can find that the temperature and heat flux profiles are very similar between NEMD with the Langevin thermostat and BTE with the thermalizing boundary. We notice that the similarity between the Langevin thermostat in NEMD and the thermalizing boundary condition in BTE is also suggested by Dunn *et al*[33], but the result is obtained with a relatively rough comparison of the temperature jump in NEMD and BTE. Here, we emphasize that not only the trends are similar, but also the temperature profile and heat flux values are quantitatively comparable. The small difference in the temperature of the sample region between BTE and NEMD results may arise from the relaxation time approximation used in the BTE calculations and the fact that the Langevin thermostat modulates the phonon density of states[50]. The comparison between NEMD simulation with the NHC thermostat and BTE with uniform heat generation also has a surprisingly good quantitative agreement, as can be seen in Fig. 7(a) and (b). The small differences of the heat flux may be due to our assumption that the same amount of energy is added to different phonon modes in BTE. The actual amount of heat added in every phonon mode in NMED is difficult to exactly obtain. Nevertheless, from the quantitatively good agreement between the NEMD and BTE results, we can make the most important conclusions for this work:

1) The NEMD simulation is comparable to phonon BTE with a proper simulation setup. The Langevin thermostat is similar to a thermalizing boundary condition in BTE. The NHC thermostat with fixed layers is similar to a uniform heat generation in the thermal reservoir with adiabatic boundaries.

2) All the non-Fourier behaviors observed in NEMD simulations, including non-linear temperature profile, temperature jump and size effect, can also be observed in the phonon BTE and have clear physical interpretations with the phonon transport picture.

To further confirm our results, we can also compare the modal temperatures from NEMD and BTE. Taking the case of $L = 13$ nm as an example, the average temperature profiles for 6 phonon branches from BTE are shown in Fig. 8. By comparing Fig. 8(a) and Fig. 4(a), we note that the thermal reservoirs can be treated as a target temperature and the temperature of different phonon bands is in equilibrium. The modal temperatures in the sample region are out of equilibrium, which is the same as in the NEMD simulations. Similarly by comparing Fig. 8(b) and Fig. 4(b), the spectral temperature profiles have the same feature that the modal temperatures are out of equilibrium not only in the sample region, but also inside the thermal reservoirs in both NEMD and BTE calculations. Note that from the BTE, a phonon branch with a longer mean free path have a smaller temperature drop throughout the domain and vice versa. A similar trend is also observed in the NEMD simulations (Fig. 4).

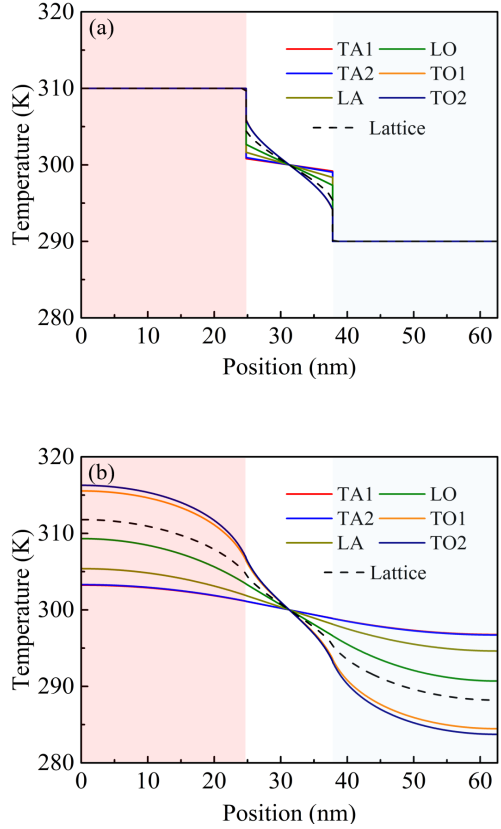

Fig. 8. (a) Averaged temperature profiles for phonon modes in 6 different branches in BTE with the thermalizing boundary conditions. (b) Temperature profiles for 6 different phonon branches in BTE with a uniform heat generation.

By the comparison between NEMD simulations and phonon BTE, we can extract the phonon interpretation of the non-Fourier heat conduction simulated by NEMD simulations. As shown in Fig. 9 (a), by using the Langevin thermostat, the thermal reservoir behaves like an infinitely large equilibrium thermal reservoir, which is similar to a blackbody in radiative heat transfer. The phonons are only emitted from the boundary of the reservoir (the small region in the reservoir that near the sample region) and all have the same temperature. All phonons enter the boundary are absorbed. By using the NHC thermostat and the VR method, as shown in Fig. 9 (b), uniform heat generation occurs in the whole thermal reservoir. The phonons are emitted into the sample region from the whole volume and continue moving inside the entire simulation domain until they scatter with each other or with the adiabatic boundary.

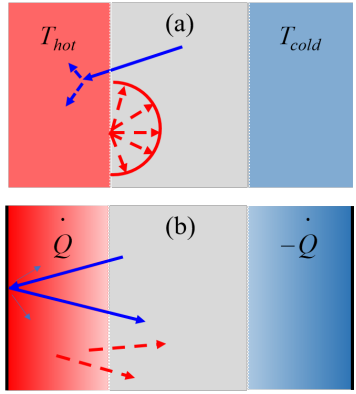

Fig. 9. Schematic illustration of the phonon interpretation of the non-Fourier heat conduction by NEMD simulations with (a) the Langevin thermostat and (b) the NHC thermostat. (a) The temperature inside thermal reservoirs remains constant ($T_{hot}$ in the heat source and $T_{cold}$ in the heat sink). The phonons are only emitted from the boundary of the reservoir (the small region in the reservoir that near the sample region) and all have the same temperature. All phonons enter the boundary are absorbed. (b) Uniform heat generation occurs in the whole thermal reservoir ($\dot{Q}$ in the heat source and $-\dot{Q}$ in the heat sink). The phonons are emitted into the sample region from the whole volume and continue moving inside the entire simulation domain until they scatter with each other or the adiabatic boundary.

## IV. Further proof by NEMD simulations

From the interpretation above, several deductions can be made. The first one is about the length of the thermal reservoirs. For the Langevin thermostat, since only the boundary (the small region in the reservoir that near the sample) influences the thermal transport, the length of the thermal reservoirs should not influence the results when it exceeds the length of the small region. For the NHC thermostat, the whole volume influences the thermal transport, so the length of the thermal reservoirs should significantly affect the results. To prove this, the temperature profiles and heat flux values in the sample region for different lengths of thermal reservoirs using the Langevin and the NHC thermostats are shown in Fig. 10. From Fig. 10(a), we see that for the Langevin thermostat, when the lengths of thermal reservoirs are larger than 0.5 nm, the length essentially has no influence on the temperature profile and heat flux. This means that only the small region near the sample influences the simulation results. This is consistent with the finding in previous studies that the length of the thermal reservoirs does not influence the results when it exceeds a critical value which depends on the time parameter of the thermostat[9,32,51]. In contrast, for the NHC thermostat in Fig. 10(b), the temperature profile and heat flux value change with the length of the thermostat. Therefore, the Langevin thermostat behaves like an infinitely large thermal reservoir while the NHC thermostat is a finite-length reservoir with uniform heat generation.

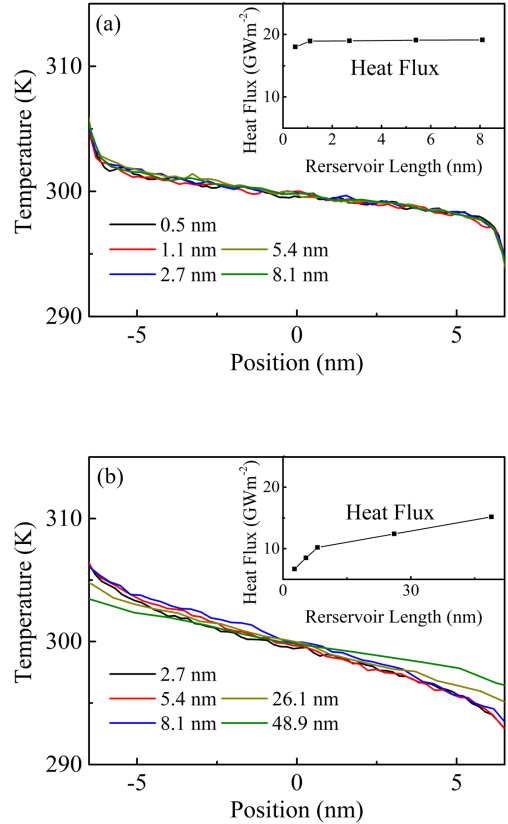

Fig. 10. Temperature profiles and heat flux values obtained by using different lengths for the thermal reservoirs. (a) The Langevin thermostat (b) the NHC thermostat.

The second one is about boundary scattering. For the infinitely large thermal reservoir, the phonons entered are absorbed, and there is no boundary scattering. For the finite-length reservoir with uniform heat generation, the phonons can continue moving inside the thermal reservoirs thus the boundary scattering should be important to the results. Thus, we set three configurations as shown in Fig. 11 to test the boundary scattering.

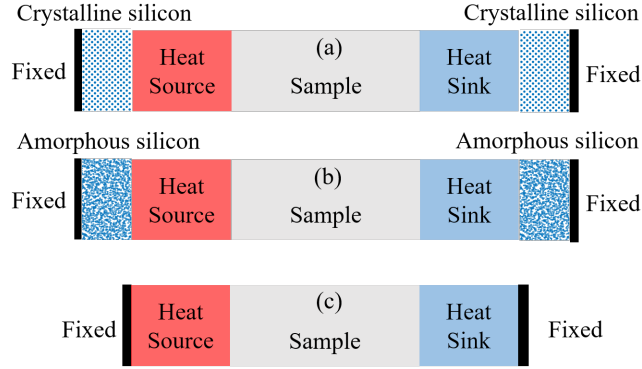

Fig. 11. Three configurations used to test boundary scattering. (a) We add 2.2 nm crystalline silicon between the thermal reservoirs and fixed layers. (b) We add 2.2 nm amorphous silicon between the thermal reservoirs and fixed layers. (c) The origin configuration used in NEMD simulation.

In the first configuration (Fig. 11(a)), we add 2.2 nm crystal silicon between the thermal reservoirs and the fixed layers. In the second configuration (Fig. 11(b)), we add 2.2 nm amorphous silicon between the thermal reservoirs and the fixed layers to generate diffuse phonon scattering. Recent wave packet simulations have clearly shown that a flat crystalline surface specularly scatters phonons while amorphous silicon induces strong diffuse scattering[52]. The third configuration is the same as in the previous cases (Fig. 11(c)). We set the length of thermal reservoirs as 8.2 nm and the length of the sample as 13 nm. The temperature and heat flux profiles are shown in Fig. 12.

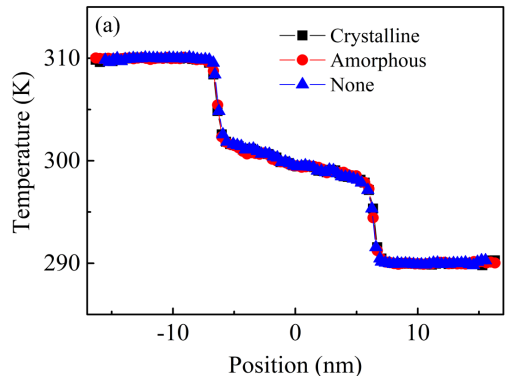

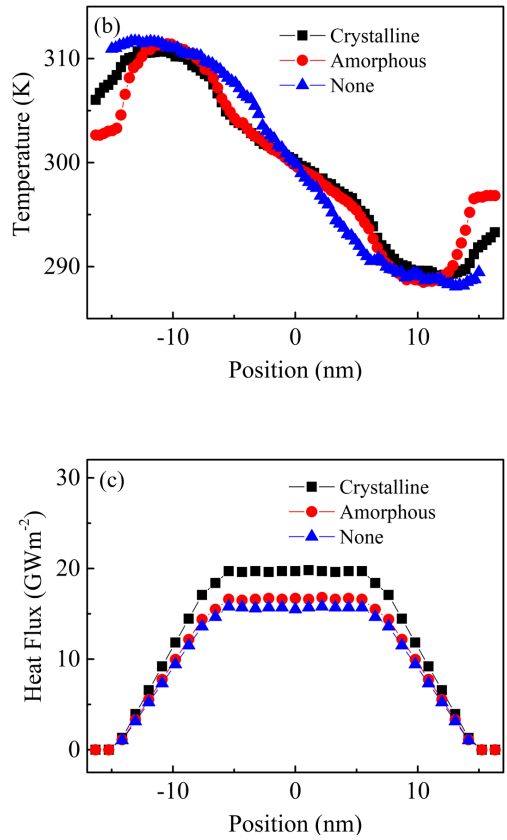

Fig. 12. Temperature profiles of (a) Langevin thermostat, (b) NHC thermostat and (c) heat flux profiles of NHC thermostat for different simulation configurations. (None) No space between fixed layers and thermal reservoirs. (Crystalline) 2.2 nm crystal silicon is added between the fixed layers and thermal reservoirs. (Amorphous) 2.2 nm amorphous silicon is added between the fixed layers and thermal reservoirs.

It can be seen in Fig. 12(a) that there is no difference between the temperature profiles of these three configurations for the Langevin thermostat. The values of the heat flux in the sample region are also the same and are not shown here for simplicity. This proves that the phonons entering the thermal reservoirs will not reach the boundary when we use the Langevin thermostat. It can be seen in Fig. 12(b) that the temperature profiles of these three configurations are quite different from each other for the NHC thermostat. The values of the heat flux in the sample region are also not the same, as shown in Fig. 12(c). This proves that the phonons entering the thermal reservoirs will reach the boundary and will not be completely absorbed in the thermal reservoirs when we use the NHC thermostat. One should also be informed that the temperature drop (or increase) near the two ends is not a simulation error. In Fig. 12(c), we have calculated the heat flux in these regions and they are proven to be zero. Since in these regions strong phonon non-equilibrium exists, the obtained temperature value is just some modal average. The drop (or increase) just indicates that the modes become more equilibrium. It does not mean there is heat flow from the heat bath to the adiabatic boundary. The origin of this phenomenon is similar to the nonlinearity of temperature and will be discussed latter.

We note that in another work by Liang *et al.,* the boundary was made into a "rough" structure to intentionally induce diffusive scattering in NEMD simulations for interface thermal resistance[50]. They also observed that the boundary roughness matters for the VR method while it is not for the Langevin thermostat. However, they believe that the VR method is more realistic while the Langevin thermostat generates artifacts. In fact, from our understanding above, the difference is merely due to the different natures of the VR method and the Langevin thermostat. As the VR method involves a volumetric heat generation, phonons can encounter the boundary. In comparison, the Langevin thermostat is an equilibrium thermostat, in which phonons are equilibrated in the reservoir before they encounter the boundary and thus the boundary atomic arrangement does not affect the phonon transport.

## V.   Discussions

Based on the phonon interpretation developed above, we can better clarify the non-Fourier phenomena observed in NEMD simulations. The nonlinearity of temperature in the sample region in NEMD simulations is neither a simulation artifact nor due to the strong scattering near the reservoirs. Instead, this nonlinearity is a physical phenomenon that is related to non-diffusive transport and local nonequilibrium of different phonon modes. In this situation, the temperature gradient not only relates to the heat flux but also relates to the local non-equilibrium of different phonon modes. This can be proved by the fact that in Fig. 12(b), we see a surprising phenomenon that there is a temperature gradient near the fixed layers even though the heat flux is zero. Therefore, in Fig. 4, the local non-equilibrium occurs in the sample for all cases, then the nonlinearity exists for all cases although the heat flux is constant. The Fourier law fails in this situation because it only describes the relationship between the temperature gradient and the heat flux and ignores the effects of the local non-equilibrium.

The abrupt temperature jump near the thermal reservoirs only appears when the Langevin thermostat is used. In the BTE framework, it is a normal phenomenon when a constant temperature boundary condition is applied and when ballistic transport appears[53]. Within the thermal reservoir, a fixed temperature is enforced, while in the sample region close to the thermal reservoir the temperature is affected by the emitted phonons from the other reservoir. Some of these emitted phonons transport ballistically and their temperature is close to the temperature of the other reservoir. This effect disappears in the diffusive regime because the phonons from the other reservoir equilibrize with other phonons during the transport process, also resulting in a continuous and linear temperature profile. In contrast, when the NHC thermostat or the VR method is used, the temperature inside the reservoirs is not enforced. Thus, the abrupt temperature jump is not obvious. The size effect is a result of ballistic transport, which is the same as the previous understanding. However, the Langevin thermostat and the NHC thermostat (or the VR method) correspond to different configurations of thermal ballistic transport, as discussed above.

The physical pictures above can also provide guidance to understand the results extracted from NEMD simulations. In the non-diffusive transport regime, we recommend using the Langevin thermostat to calculate the thermal conductance (or the apparent thermal conductivity) of the sample. The conductance should be obtained using $C = q/\Delta T$, where $q$ is the heat flux value and $\Delta T$ is the temperature difference of the thermal reservoirs. The conductance obtained using the Langevin thermostat is similar to the case when a sample is coupled to two infinite thermal reservoirs, which has been widely adopted in the BTE[53] or Landauer framework[54]. By using this method, the results by NEMD can be comparable to those by other simulation methods including BTE, AGF and HNEMD[32]. Moreover, the finite-size effect of thermal conductance (or the apparent thermal conductivity) can be described by the analytical model derived by BTE[55]. Using the NHC thermostat or the VR method is still reasonable if the sample size is much larger than the mean free path. Nevertheless, in the non-diffusive regime, the thermal conductance obtained using either temperature difference of the sample boundary or the temperature difference of the thermal reservoir, will be dependent on the size of the reservoirs and the boundary atom arrangement. Also, because of the non-equilibrium of the thermal reservoir, it is difficult to clearly identify the physical meaning of the obtained results.

Our results can also guide the measurement and control of thermal transport in real solid-state devices. We have shown that the nanoscale phonon transport characteristics including the temperature profile, heat flux value and modal temperature strongly depend on the applied thermal reservoirs in NEMD simulations. This is also true for real devices. The Langevin thermostat, i.e. an infinitely large equilibrium thermal reservoir can be realized when a dielectric film is sandwiched between two metallic films. Since the electron-phonon mean free path in the metals is much smaller than the phonon mean free path in the dielectric film, the two interfaces between the dielectric film and the metallic film can be assumed to be equilibrium thermal reservoirs at the fixed temperature[43]. The abrupt temperature jump exists in the two interfaces. The apparent thermal conductivity can be defined in the same way as for the Langevin thermostat. In contrast, the NHC thermostat or the VR method, i.e. a uniform heat generation in the finite thermal reservoir provides large non-equilibrium outlets. Practical heating techniques, such as optical heating or electrical heating, can result in this large non-equilibrium reservoirs which selectively heat the optical phonon modes[56]. By using these techniques, the length of the thermal reservoir and the boundary conditions can significantly influence the relationship between the response and the perturbation such as the heat flux and the temperature profile.

## VI. Summary and conclusions

To summarize, we performed NEMD simulations and mode-resolved phonon BTE using different reservoirs to understand the nanoscale non-Fourier phonon transport. A quantitative agreement between the NEMD and phonon BTE is achieved, which shows that the thermal transport heavily depends on the reservoirs. The Langevin thermostat behaves like an infinitely large equilibrium thermal reservoir, which is similar to a black surface in radiative heat transfer. Therefore, the phonons are only emitted from the boundary of the reservoir (the small region near the sample region) and all have the same temperature. All phonons enter this boundary are absorbed. The NHC thermostat and the VR method behave like finite-size non-equilibrium phonon source/sink with uniform energy deposition/extraction. Therefore, the phonons are emitted into the sample region from the whole volume and continue moving until they scatter with each other or the adiabatic boundary. All the non-Fourier behaviors observed in NEMD simulations, including the nonlinearity of the temperature profile, the abrupt temperature jump in some situations, and the length-dependent thermal conductivity, have clear physical interpretations. They are due to the combination of non-diffusive phonon transport and non-equilibrium among different phonon modes. These results not only help to understand the non-Fourier heat conduction in NEMD simulations, but also provide better understanding of the thermal transport in confined nanostructures.

**Acknowledgements**

Y.H. and H.B. acknowledge the support by the National Natural Science Foundation of China (Grant No. 51676121). X.G. acknowledges the support by the National Natural Science Foundation of China (Grant No. 51706134). Z.F. acknowledges the support from the National Natural Science Foundations of China (Grant No. 11974059). Simulations were performed at center for High Performance Computing (π) of Shanghai Jiao Tong University. The authors acknowledge the assistance of Xinyu Zhang in the computation process.